\documentclass[preprint,showpacs,preprintnumbers,amsmath,amssymb]{revtex4-1}

\usepackage{graphicx}
\usepackage{epstopdf}

\usepackage{dcolumn}
\usepackage{bm}
\usepackage{color}

\usepackage{feynmf}
\usepackage{slashed}
\usepackage{enumerate}

\usepackage[scientific-notation=true]{siunitx} 

\usepackage[utf8]{inputenc}

\usepackage[colorlinks=true,urlcolor=blue,anchorcolor=blue,citecolor=blue,filecolor=blue,linkcolor=blue,menucolor=blue,linktocpage=true]{hyperref} 

\usepackage{appendix}
\usepackage{amsmath,amsthm,amssymb,mathrsfs}
\usepackage{bbold}
\usepackage{bbm}

\begin{document}

\title{Muon precession from the aspect of Dirac equations}

\author{Jinbo He}

\author{Lei Ming}
\thanks{minglei@mail.sysu.edu.cn}

\author{Yi-Lei Tang}
\thanks{tangylei@mail.sysu.edu.cn}

\author{Qiankang Wang}
\thanks{wangqk6@mail2.sysu.edu.cn}

\author{Hong-Hao Zhang}
\affiliation{School of Physics, Sun Yat-Sen University, Guangzhou 510275, China}


\begin{abstract}
In this paper, we would like to compute the muon anomalous precession frequency through solving the wave functions of the Dirac equations straightforwardly. The precession of a (anti-)muon with an anomalous magnetic momentum term is calculated together with the quantum corrections. Lorentz violation terms up to the lowest non-trivial order is introduced, and their effects on anomalous precession are evaluated perturbatively.
\end{abstract}
\pacs{}

\keywords{}

\maketitle
\section{Introduction}

The precession experiments are critical to determine the gyro-magnetic ratio of a massive particle\cite{Muong-2:2004fok, Muong-2:2021vma, Muong-2:2021ojo, Muong-2:2023cdq}. The Dirac equation predicts that the gyro-magnetic ratio of an elementary charged particle $g_X$ should be $2$, although subdominant corrections might arise through the self-energy loop diagrams. This offers us chances to probe the possible effects contributed from the beyond standard model (BSM) particles in the loops by comparing the measured $g_X-2$ with the theoretical predictions.

A practical precession experiment for the measurement of the $g_{\mu}-2$ involves injecting a bunch of polarized relativistic (anti-)muons into a cyclotron-like equipment (or sometimes a ``Penning trap'') configured by a combination of a vertical magnetic field and an electric field. Then these (anti-)muons circulates while decaying into electrons(positrons) detected by the detectors deployed nearby. As the polarization direction of the muons precess, the main directions of the decay products evolve periodically. The precise $g_{\mu}-2$ is then extracted from these temporal information.

For comparison with the theoretical evaluations, one has to calculate the precession of a relativistic (anti-)muon with an anomalous magnetic momentum circulating in the equipment. In the literature, the famous Thomas-Bargmann-Michel-Telegdi, or Thomas-BMT formula\cite{Thomas:1927yu, Bargmann:1959gz, Jackson:1998nia, Berestetskii:1982qgu} was achieved by assuming the muons to be point-like particles with a particular spin direction, and to extend the classical polarization vector into a relativistic four-dimensional vector to solve the classical equation of motions. In this case, the muons are regarded as classical objects with classical spin directions. Such a scenario might be viable since the size of the muon's macroscopic trajectories are much larger than the wave packets of the muon particles. As a fundamentally well-known ``quantum effect'', whether there will be a sheer alternative ``quantum understanding'' of such a.phenomenon is particularly interesting.

Sometimes, exotic interactions between muons and photons arise. For an example, Lorentz violation terms might affect the precession periods to correct the measured $g_{\mu}-2$ value\cite{Bluhm:1999dx, Bluhm:1997qb, Bluhm:1997ci, Kostelecky:2013rta, Gomes:2014kaa, Aghababaei:2017bei}. In the literature, such terms are manipulated by the non-linear Foldy-Wouthuysen transformation\cite{Bluhm:1999dx, Bluhm:1997qb, Bluhm:1997ci} and finally contribute to an effective magnetic anomalous term as expanded to the lowest non-trivial order, therefore reducing to the scenario that a classical point-like particle moves and precesses within the equipment. Recent extractions of the Lorentz violation parameter data can be found in Ref.~\cite{Muong-2:2007ofc, Quinn:2019ppv, Mitra:2023imh}.

In contrast to the literature, we aim at resolving the Dirac equations for a straightforward solution to the muon precession processes in a uniform magnetic field. In this paper, the muons are regarded as ``wave-packets'' rather than the classical ``point-like classical objects''. The results are fundamentally based upon ``quantum mechanics'', and in some cases, tiny quantum corrections arise, although these corrections are usually too small to accommodate the recent measured $g_{\mu}-2$ anomaly.

When the Lorentz violation terms are included, usually no precise analytic solution exists. In this case, the perturbation method can be utilized. Up to the lowest order, our calculations are compatible with the estimations through the Foldy-Wouthuysen transformation. In principle, our strategy can be extended up to higher orders.

In the following of this paper, we will describe the basic equations and concepts about how we extract the precession informations from a solution to the (modified-) Dirac equations. At the same time, the simpler Dirac equations with a usual anomalous magnetic term are solved. Then Lorentz violation terms are introduced and the equations are solved perturbatively up to the first order. Finally, we summarize this paper.

\section{Lagrangian and basic concepts}

Let us start with the effective Lagrangian of a muon particle. Besides the usual term $\overline{\psi}\frac{e(g-2)}{4m}\sigma^{\mu\nu}\partial_{\nu}A_{\mu}\psi$ which contributes to the anomalous magnetic momentum, we shall consider the effect of Lorentz violation terms. The effective Lagrangian up to the lowest orders is
\begin{equation}\label{action}
	\begin{aligned}
		\mathcal{L} &=i\overline{\psi}\gamma^{\mu}D_{\mu}\psi+\frac{i}{2}c_{\mu\nu}\overline{\psi}\gamma^{\mu}D^{\nu}\psi+\frac{i}{2}d_{\mu\nu}\overline{\psi}\gamma_{5}\gamma^{\mu}D^{\nu}\psi\\
		&+m\overline{\psi}\psi+\overline{\psi}\frac{e(g-2)}{4m}\sigma^{\mu\nu}\partial_{\nu}\mathcal{A}_{\mu}\psi-a_{\mu}\overline{\psi}\gamma^{\mu}\psi-b_{\mu}\overline{\psi}\gamma_{5}\gamma^{\mu}\psi-\frac{1}{2}H_{\mu\nu}\overline{\psi}\sigma^{\mu\nu}\psi,
	\end{aligned}
\end{equation}
in which $D_\mu=\partial_\mu-i e \mathcal{A}_\mu$ is the covariant derivative,  $\mathcal{A}_\mu$ the vector potential, $g$ the Land\'{e} $g$-factor, $\sigma^{\mu\nu}=\frac{i}{2}[\gamma^{\mu},\gamma^{\nu}]$, $\gamma^\mu$ the Dirac matrices in the Pauli-Dirac representation,
\begin{equation*}
    \gamma_0=\left(
        \begin{array}{cc}
             \mathbb{1}&0\\
             0&-\mathbb{1}
        \end{array}
    \right),\\
    \gamma_{i}= \left(
        \begin{array}{cc}
            0&\sigma_{i}\\
            \sigma_{i}&0
        \end{array}
    \right),\ \
    \gamma_{5}= \left(
        \begin{array}{cc}
            0&\mathbb{1}\\
            \mathbb{1}&0
        \end{array}
    \right)
\end{equation*}
with $\sigma_i$ being the Pauli matrices. $a_\mu$, $b_\mu$, $c_{\mu\nu}$, $d_{\mu\nu}$ and $H_{\mu\nu}$ are the effective couplings that could arise from the extended theories of Standard Model with higher energy scale -- probably near Planck mass $m_{\rm pl}~10^{19}$ GeV.
In this paper, we focus on the effective Lagrangian without concerning their sources.

Variation of (\ref{action}) with respect to the $\psi$ field gives the modified Dirac equation,
\begin{equation}\label{eqDirac}
    \begin{aligned}
    &i\gamma^{\mu}\partial_{\mu}\psi+\frac{i}{2}c_{\mu\nu}\gamma^{\mu}\partial^{\nu}\psi-\frac{i}{2}d_{\mu\nu}\gamma^{\mu}\gamma_{5}\partial^{\nu}\psi
    +e\gamma^{\mu}\mathcal{A}_{\mu}\psi+m\psi-a_{\mu}\gamma^{\mu}\psi\\
    &+b_{\mu}\gamma^{\mu}\gamma_{5}\psi-\frac{1}{2}H_{\mu\nu}\sigma^{\mu\nu}\psi+\frac{e}{2}c_{\mu\nu}\gamma^{\mu}\mathcal{A}^{\nu}\psi-\frac{e}{2}d_{\mu\nu}\gamma^{\mu}\gamma_{5}\mathcal{A}^{\nu}\psi+\frac{e(g-2)}{4m}\sigma^{\mu\nu}\partial_{\nu}\mathcal{A}_{\mu}\psi=0.
    \end{aligned}
\end{equation}

\begin{figure}
    \centering
    \includegraphics[width=0.8\textwidth]{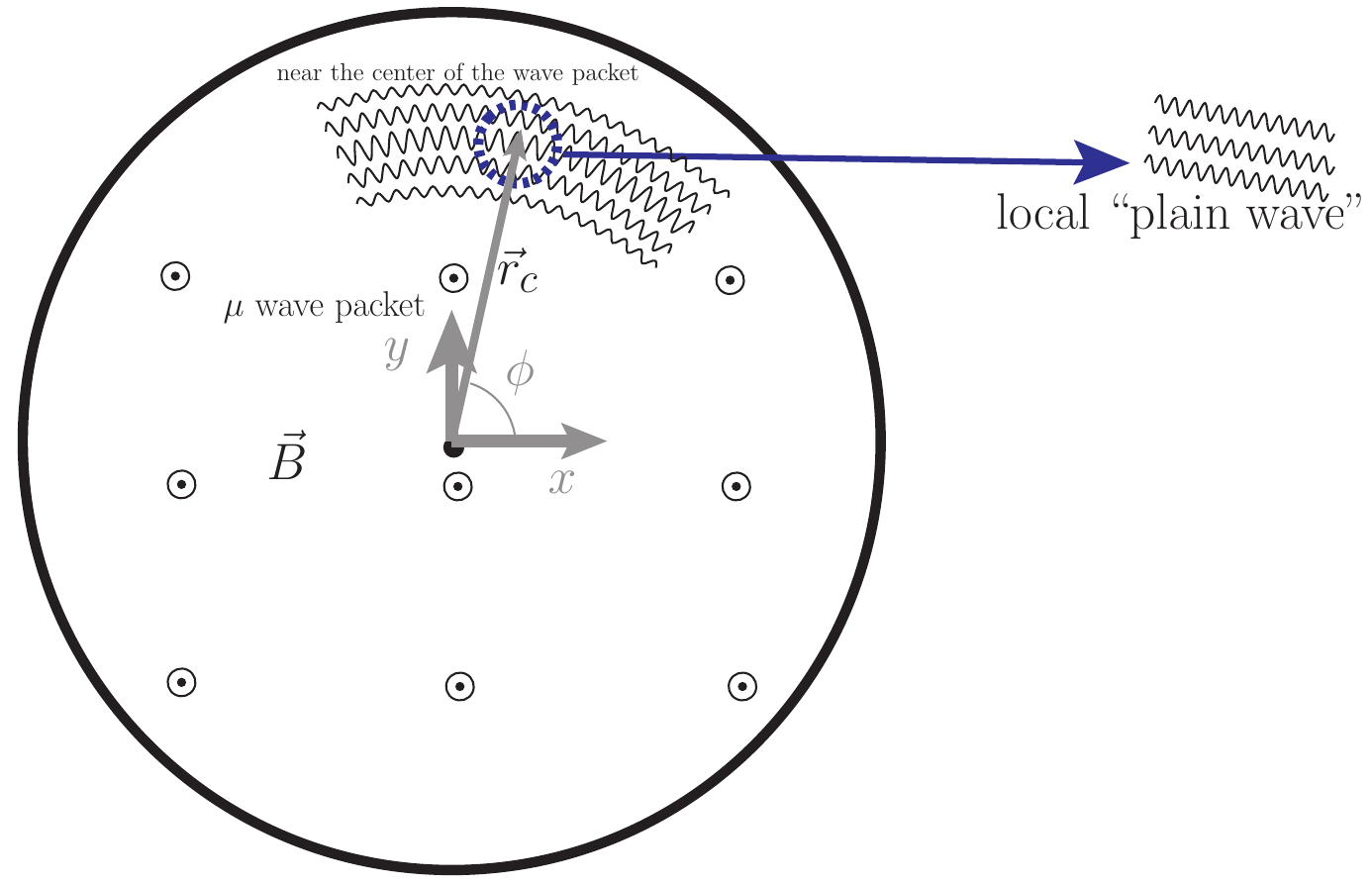}
    \caption{Sketched picture of a wave packet of a muon circulating horizontally with in the range of a vertical uniform magnetic field. Zooming up the center of the wave packet gives something similar to a ``plain wave'' in the local area.}
    \label{CyclotronSketch}
\end{figure}

The following of this paper aims at solving this Dirac equation. Before doing this, let us depict a simplified picture of a muon wave-packet circulating within a uniform magnetic field, as shown in Fig.~\ref{CyclotronSketch}.
The uniform magnetic field is along the $z$-axis as in Fig.~\ref{CyclotronSketch} so that $\vec{B}=B\Vec{z}$. In this paper, we adopt the symmetric gauge of the vector potential $\mathcal{A}_{\mu}$ which can be expressed as
\begin{equation} \label{VectorPotential}
\mathcal{A}_\mu=\left(
\begin{array}{c}
	0\\
	\frac{1}{2}By\\
	-\frac{1}{2}Bx\\
	0
	 \end{array}
\right).
\end{equation}
The wave function of a wave packet can be parameterized to be
\begin{equation}
\psi\equiv \left( \begin{array}{c}
  \psi_1 \\
  \psi_2 \\ 
  \psi_3 \\ 
  \psi_4
  \end{array}\right),~\psi_i(r, \phi) \simeq \left[A_{i}(\phi, t) + B_{i}(\phi, t)(r-r_{c})^{2} + \mathcal{O}(r-r_c)^3\right]e^{in_{i}\phi}, \label{GeneralPackets}
\end{equation}
where we rely on cylindrical coordinates so that $r=\sqrt{x^2+y^2}$, $x = r \cos\phi$, and $y= r \sin\phi$. The $z$-dependence is neglected since its evolution does not contribute to the precession processes. $r_c$ is defined to be the distance between the center of the wave packet and the circulating center. The $(r-r_c)$ term is neglected for the consideration that near the center of the wave packet, radial momentum is expected to be zero as the particle is moving along a circle. Truncating the series up to $(r-r_c)^2$ order gives the $A_i(\phi)$ and $B_i(\phi)$ in (\ref{GeneralPackets}), which are assumed to be ``flat'' enough within the range of the wave packet, and accommodate sufficient number of complete wavelengths. Therefore, just as the standard procedure when solving the Schr\"odinger equations of the one-dimensional scattering problems, we can neglect the $t$ and $\phi$ dependence of $A_i$ and $B_i$ within the wave packet to acquire a ``stationary state''. The real global stationary state wave function should satisfy the periodic boundary condition for its unique value at each point, however since this time we confine ourselves inside the wave packet, the periodic boundary condition $\psi_i(r, \phi)=\psi_i(r, \phi+2 \pi)$ is abandoned. That is to say, we only discuss the wave functions
\begin{equation}
\psi_i(r, \phi) \simeq \left[A_{i} + B_{i}(r-r_{c})^{2} + \mathcal{O}(r-r_c)^3\right]e^{in_{i}\phi}, \label{GeneralPackets_Stationary}
\end{equation}
where $A_i$ and $B_i$ are constants.

For the convenience of a clearer discussion, before solving the Dirac equations officially, we in advance give the assertion that usually the values of the $n_i$'s are extremely close to each other. Actually, we will see in (\ref{n1234}) that $n_{2}=n_{3}+1$ and $n_{4}=n_{1}+1$. Therefore, if we zoom inside the wave packet, one acquires a local ``plane wave'' near the center of the wave packet. After extracting a common factor $e^{i n_1 \phi}$, we have
\begin{equation}
\psi \simeq e^{i n_1 \phi} \left( \begin{array}{c}
  A_1 \\
  A_2 e^{i (n_3-n_1+1) \phi} \\ 
  A_3 e^{i (n_3-n_1) \phi} \\ 
  A_4 e^{i  \phi}
  \end{array}\right). \label{GeneralPackets_Plain}
\end{equation}
The factor $e^{i n_1 \phi}$ can be reduced to the ``local plain-wave'' form of $\approx e^{i \frac{\vec{p}}{2} \cdot (\vec{r}-\vec{r_c})} e^{i n_1 \phi_c}$ where the $\vec{p}$ replaces the $2 \frac{n_1}{r_c} (-\sin \phi, \cos \phi, 0)$, and is perpendicular to the $\phi$-direction, which can be expressed as the local ``physical momentum vector'' of the local plain wave. The exotic factor $2$ originate from the difference between the physical momentum and the canonical momentum. The detailed strict discussions will be addressed later. Therefore, the remained row vector part in the (\ref{GeneralPackets_Plain}) contains the polarization information, and becomes crucial for the precession discussions.

As the time elapses, the overall envelope of the wave packet moves forward while the ``stationary wave functions'' inside the envelope remains unchanged except a universal $e^{i E t}$ oscillating factor. Therefore, the $\phi$-dependence of (\ref{GeneralPackets_Plain}) can be treated as the evolution of polarization as time changes.

In principle one can decompose (\ref{GeneralPackets_Plain}) into a cumbersome combination of a group of spin eigenstates to extract the precession frequency. In this paper, we adopt an alternative trick. Notice that the common precession experiments aim at measuring the difference $\omega_a = \omega_s - \omega_c$ between the spin precession frequency $\omega_s$ and the so-called ``cyclotron frequency'' $\omega_c = \frac{e B}{m_{\mu}}$, which means, one is actually comparing the spin direction with the particle's momentum direction and observing the evolution of such a difference. Then, it is easy to rely on a instant reference that the momentum direction of the wave packet is fixed along the y-axis. For the positive energy solution indicating the $\mu^-$ particle, the particle moves anti-clockwise at Fig.~\ref{CyclotronSketch}, so this can be done by rotating the experiment reference frame defined in Fig.~\ref{CyclotronSketch} by an angle of $\phi$ (anti-clockwise) along the z-axis, and (\ref{GeneralPackets_Plain}) becomes
\begin{equation}
    e^{-i \phi S_3} \psi \overset{\sim}{\propto} \left( \begin{array}{c}
  A_1 \\
  A_2 e^{i (n_3-n_1) \phi} \\ 
  A_3 e^{i (n_3-n_1) \phi} \\ 
  A_4
\end{array}\right), \label{GeneralPackets_Rotated}
\end{equation}
where the $\overset{\sim}{\propto}$ indicates that such a ``proportional to'' symbol includes an operation of approximation.

If $n_1=n_3$, (\ref{GeneralPackets_Rotated}) clearly prompts that the spins of the muon relative to its momentum directions remain unchanged as the momentum direction evolves, which means $\omega_s=\omega_c$ and no anomalous precession arises. Apparently nonzero $n_3-n_1$ induces anomalous precession processes, and the spin vector restores after the wave packet rotates by an angle of $\frac{2 \pi}{n_3 - n_1}$, inducing a anomalous precession frequency
\begin{equation}\label{omega1}
    \omega_a = (n_3 - n_1) \omega_c,
\end{equation}
for positive energy solutions indicating the particles ($\mu^-$), and the wave packet moves anti-clockwise. Similar discussions show
\begin{equation}\label{omega2}
    \omega_a = (n_1 - n_3) \omega_c,
\end{equation}. 
for negative energy solutions indicating the anti-particles ($\mu^+$) moving clockwise. Both (\ref{omega1}) and (\ref{omega2}) are the fundamental equations we rely throughout this paper.

In the rest part of this paper, we at first consider the contribution only from the usual anomalous magnetic moment term, i.e., the term containing $\partial_\nu A_\mu$ in (\ref{action}), and then estimates perturbatively the solutions with the existence of each effective Lorentz violating terms  while keeping the others zero.

\section{Contribution of the anomalous magnetic moment term}	
We now take into account only the contribution of the anomalous magnetic moment in (\ref{action}), which is described by an extra potential term proportional to $g-2$ in (\ref{eqDirac}),
\begin{equation}
	i\gamma^{\mu}\partial_{\mu}\psi+\left(m+e\gamma^{\mu}\mathcal{A}_{\mu}+e\frac{g-2}{4m}\sigma^{\mu\nu}\partial_{\nu}\mathcal{A}_{\mu}\right)\psi=0.
\end{equation}

By the coordinate transformation between the Cartesian corrdinate system and the cylindrical one, 
\begin{equation}
	\frac{\partial }{\partial x}=\cos\phi\frac{\partial}{\partial r}-\frac{\sin\phi}{r}\frac{\partial}{\partial \phi},\ \ \frac{\partial }{\partial y}=\sin\phi \frac{\partial }{\partial r}+\frac{\cos\phi}{r}\frac{\partial}{\partial \phi},
\end{equation}
one is then able to re-write the Dirac equation as\cite{Old1928}
\begin{gather}
	(E+m+K)\psi_{1}+ie^{-i\phi}\left(\frac{\partial}{\partial r}-\frac{i}{r}\frac{\partial}{\partial \phi}+\frac{e B r}{2}\right)\psi_{4}=0, \label{Dirac1}\\
	-ie^{i\phi}\left(\frac{\partial}{\partial r}+\frac{i}{r}\frac{\partial}{\partial \phi}-\frac{e B r}{2}\right)\psi_{1}+(-E+m-K)\psi_{4}=0, \label{Dirac2}\\
	(E+m-K)\psi_{2}+ie^{i\phi}\left(\frac{\partial}{\partial r}+\frac{i}{r}\frac{\partial}{\partial \phi}-\frac{e B r}{2}\right)\psi_{3}=0, \label{Dirac3}\\
	-ie^{-i\phi}\left(\frac{\partial}{\partial r}-\frac{i}{r}\frac{\partial}{\partial \phi}+\frac{e B r}{2}\right)\psi_{2} +(-E+m+K)\psi_{3}=0, \label{Dirac4}
\end{gather}
where $K\equiv\frac{ (g-2)eB}{4m}$ and $\psi\equiv\left(\psi_1, \psi_2, \psi_3, \psi_4\right)^\mathrm{T}$.

To solve the above equations, we adopt the expansion (\ref{GeneralPackets_Stationary}). It is then easy to find 
\begin{equation}\label{n1234}
   n_{2}=n_{3}+1 \quad {\rm and} \quad  n_{4}=n_{1}+1
\end{equation}
by comparing the phase of each term. Keeping only the leading  $(r-r_c)^0$, the equations are reduced into 
\begin{equation}\label{eq38}
	\left(
	\begin{array}{cc}
		E+m-K&-i\left(\frac{n_{3}}{r_{c}}+\frac{e B r_{c}}{2}\right)\\
		-i\left(\frac{n_{2}}{r_{c}}+\frac{e B}{2}r_{c}\right)&-E+m+K
	\end{array}
	\right)\left(
	\begin{array}{c}
		A_{2}\\
		A_{3}
	\end{array}
	\right)=0
\end{equation}
and
\begin{equation}\label{eq39}
	\left(
	\begin{array}{cc}
		E+m+K&i\left(\frac{n_{4}}{r_{c}}+\frac{e B r_{c}}{2}\right)\\
		i\left(\frac{n_{1}}{r_{c}}+\frac{e B}{2}r_{c}\right)&-E+m-K
	\end{array}
	\right)\left(
	\begin{array}{c}
		A_{1}\\
		A_{4}
	\end{array}
	\right)=0,
\end{equation}
from which one has
\begin{equation}\label{eq40}
	E^{2}-m^{2}=2KE-K^{2}+\frac{n_{3}(n_{3}+1)}{r_{c}^{2}}+\frac{e B}{2}(2n_{3}+1)+\frac{e^{2}B^{2}r_{c}^{2}}{4}
 \end{equation}
 \begin{equation}\label{eq41}
     E^{2}-m^{2}=-2KE-K^{2}+\frac{n_{1}(n_{1}+1)}{r_{c}^{2}}+\frac{e B}{2}(2n_{1}+1)+\frac{e^{2}B^{2}r_{c}^{2}}{4}.
 \end{equation}
 
For the equations to be consistent, we are led to the constraint
\begin{equation}
	4KE+\frac{n_{3}(n_{3}+1)-n_{1}(n_{1}+1)}{r_{c}^{2}}+e B (n_{3}-n_{1})=0,
\end{equation}
the variable $n_3-n_1$ can then be evaluated. Another method is to solve (\ref{eq40}) and (\ref{eq41}) directly. Together with (\ref{n1234}) these can be solved to get 
\begin{gather}
    n_{1}=\frac{1}{2}\left(-1-e B r_{c}^{2}\pm \Delta\right),\ n_{4}=\frac{1}{2}\left(1-e B r_{c}^{2}\pm \Delta\right),\label{solution1}\\
    n_{3}=\frac{1}{2}\left(-1-e B r_{c}^{2}\pm \Delta'\right),\ n_{2}=\frac{1}{2}\left(1-e B r_{c}^{2}\pm \Delta'\right), \label{solution2}
\end{gather}
where 
\begin{gather}    \Delta=\sqrt{1+4E^{2}r_{c}^{2}+8EKr_{c}^{2}+4K^{2}r_{c}^{2}-4m^{2}r_{c}^{2}}\simeq \sqrt{1+4E^{2}r_{c}^{2}+8EKr_{c}^{2}-4m^{2}r_{c}^{2}},\\
    \Delta'=\sqrt{1+4E^{2}r_{c}^{2}-8EKr_{c}^{2}+4K^{2}r_{c}^{2}-4m^{2}r_{c}^{2}}\simeq \sqrt{1+4E^{2}r_{c}^{2}-8EKr_{c}^{2}-4m^{2}r_{c}^{2}}.
\end{gather}
From (\ref{eq38}) and (\ref{eq39}), the relations between $A_{1,2,3,4}$ becomes 
\begin{gather}
   \frac{A_{4}}{A_{1}}=\frac{i\left(\frac{n_{1}}{r_{c}}+\frac{p}{2}\right)}{E+K-m},\ \frac{A_{1}}{A_{4}}=\frac{-i\left(\frac{n_{4}}{r_{c}}+\frac{p}{2}\right)}{E+K+m},\label{A4A1}\\
    \frac{A_{3}}{A_{2}}=\frac{i\left(\frac{n_{2}}{r_{c}}+\frac{p}{2}\right)}{-E+m+K},\ \frac{A_{2}}{A_{3}}=\frac{i\left(\frac{n_{3}}{r_{c}}+\frac{p}{2}\right)}{E-K+m}.\label{A3A2}
\end{gather}
Then we have to determine which solution in (\ref{solution1}) and (\ref{solution2}) should be adopted. Compute the $(r-r_c)^1$ order contributions of (\ref{Dirac1})-(\ref{Dirac4}), and acquire
\begin{eqnarray}
    B_{4}&=&\left(-\frac{e B}{4}+\frac{n_{4}}{2r_{c}^{2}}\right)A_{4},\\
    B_{1}&=&\left(\frac{e B}{4}-\frac{n_{1}}{2r_{c}^{2}}\right)A_{1},\\
    B_{3}&=&\left(\frac{e B}{4}-\frac{n_{3}}{2r_{c}^{2}}\right)A_{3},\\
    B_{2}&=&\left(-\frac{e B}{4}+\frac{n_{2}}{2r_{c}^{2}}\right)A_{2}.
\end{eqnarray}
One might expect all $B_i/A_i<0$ for $i=1,2,3,4$ for a wave packet should peak, rather than ``valley'' at $r=r_c$ according to (\ref{GeneralPackets_Stationary}). It is impossible for all the components $A_i + B_i (r-r_c)^2$ to satisfy this condition because $n_4 > n_1$, so it is easy to verify that $\frac{B_4}{A_4} \frac{B_1}{A_1} \leq 0$. Therefore if $A_1 + B_1 (r-r_c)^2$ peaks, then $A_2 + B_2 (r-r_c)^2$ must ``valley''. To solve this problem completely requires time-dependent wave functions in (\ref{GeneralPackets}) and calculate the evolution of the wave packet envelopes, which is beyond our discussions of precession processes. Another practical concern is that we only concede to require $\psi^+ \psi \approx \sum\limits_{i=1,2,3,4} [A_i A_i^* + (A_i^* B_i + B_i^* A_i)(r-r_c)^2)]$ peaks at $r=r_c$. This is easy to accomplish and we only point out that we do not have to determine $n_{1,2,3,4}$ very precisely. Further more, we would like the wave functions around $r \approx r_c$ to be as ``flat'' as possible to achieve the validity of a local ``plain wave'' as described in (\ref{GeneralPackets_Plain}). Since all $n_{1,2,3,4}$ are particularly close to each other, the ``flatness'' condition requires
\begin{equation}
    \left|\frac{2 n_i}{e B r_c^2} - 1\right| \ll 1
\end{equation}
Define $p=e B r_c$ as the ``classical physical momentum'' of a circulating muon, as we have already mentioned in advance shortly after (\ref{GeneralPackets_Plain}),
\begin{equation}
    \frac{2 n_i}{r_c} \simeq p. \label{pDef}
\end{equation}
We note again that the factor of $2$ originate from the difference between the definitions of the physical momentum and the canonical momentum. In the following of this paper, we sometimes substitute the $n_i$ with $r_c p/2$ directly as a very good approximation. Therefore, from (\ref{pDef}) we know that $n_{1,2,3,4} \gg 1$ so that only  ``+'' can be adopted in (\ref{solution1}) and (\ref{solution2}). Other solutions are abandoned. 
Thus,
\begin{equation}\label{n3-n1}
	\begin{aligned}
		n_{3}-n_{1}&=\frac{r^{2}_{c}}{2}\left[\sqrt{\frac{1}{r_{c}^{4}}+\frac{4(E^{2}-2EK-m^{2})}{r^{2}}}-\sqrt{\frac{1}{r_{c}^{4}}+\frac{4(E^{2}+2 E K-m^{2})}{r^{2}}}\right]\\
		&\simeq -\frac{4 E r_{c}^{2}K}{\sqrt{1+4(E^{2}-m^{2})r_{c}^{2}}} \simeq -\frac{2 E p^{2}a_{\mu}}{m\sqrt{e^{2}B^{2}+4p^{4}}},
	\end{aligned}
\end{equation}
with $a_\mu=(g-2)/2$. Here, we have adopted from (\ref{solution1}) and (\ref{solution2}) so that 
\begin{equation}
    E = \pm \frac{1}{2}\left( \sqrt{m^2 + (\frac{n_3}{r_c} + \frac{p}{2})(\frac{n_2}{r_c} + \frac{p}{2})} + \sqrt{m^2 + (\frac{n_1}{r_c} + \frac{p}{2})(\frac{n_4}{r_c} + \frac{p}{2})} \right) \simeq \pm \sqrt{p^2+m^2},
\end{equation}
which is similar to the usual energy/momentum relations of a point-like classical particle. Therefore, according to (\ref{omega1}) and (\ref{omega2}), the precession frequency $\omega_a$ of muon in magnetic field is
\begin{equation}\label{omegap}
	\omega_a\simeq\left\lbrace \begin{array}{l} (n_{3}-n_{1})\frac{eB}{|E|}\text{, for $E>0$} \\ (n_{1}-n_{3})\frac{eB}{|E|}\text{, for $E<0$} \end{array} \right. =-\frac{2eBp^{2}a_{\mu}}{m\sqrt{e^{2}B^{2}+4p^{4}}}
\end{equation}
This result is universal for both muons and anti-muons since it is the same for both positive and negative energy solutions.

Compared with the familiar Thomas-BMT results\cite{Thomas:1927yu, Bargmann:1959gz, Jackson:1998nia, Berestetskii:1982qgu}, (\ref{omegap}) contains more ``quantum effects'' and seems to be quite different. Notice that the uncertainty principle requires that
\begin{equation}
    p r_c \approx p^{2}/eB \gg 1,
\end{equation}
which means that the cyclotron radius, or equivalently in order of quantity, the circumference, should accommodate sufficient number of wavelengths for the validity that the circulating muon looks like a classical point-like particle. therefore 
\begin{equation}
    \omega_{a, c} \simeq -\frac{e B a_{\mu}}{m}
\end{equation}
which is compatible with the Thomas-BMT results. Moreover, one can estimate the error induced by quantum effects
\begin{equation}\label{QuantumCorrections}
	R=\frac{\omega_{a, c}-\omega_{a}}{\omega_{a, c}}\simeq -\frac{e^{2}B^{2}}{8 p^{4}} \hbar^2.
\end{equation}
We take $B=1.45 {\rm T}$ and $p=1.65\times 10^{-18} {\rm kg\cdot m\cdot s^{-1}}$ as some practical values referenced from Ref.~\cite{Muong-2:2004fok, Muong-2:2021vma, Muong-2:2021ojo, Muong-2:2023cdq}, we get $R\simeq 10^{-35}$.



\section{contribution of $d_{\mu\nu}$ term as an example}
Now we are going to include the Lorentz violation contributions separately. In this section, we adopt $d_{\mu \nu}$ as a paradigm to present how these terms perturbatively distort the wave functions and finally affect the $\omega_a$ values in detail. Other terms share similar cumbersome processes and we will later give only the sketched processes of some key expressions with the results. For the $d_{\mu \nu}$ contributions, the Dirac equation is obtained to be
\begin{equation}
    i\gamma^{\mu}\partial_{\mu}\psi+e\gamma^{\mu}\mathcal{A}_{\mu}\psi-\frac{i}{2}d_{\mu\nu}\gamma^{\mu}\gamma_{5}\partial^{\nu}\psi-\frac{e}{2}d_{\mu\nu}\gamma^{\mu}\gamma_{5}\mathcal{A}^{\nu}\psi+m\psi+\frac{e(g-2)}{4m}\sigma^{\mu\nu}\partial_{\nu}\mathcal{A}_{\mu}\psi=0.
\end{equation}

Note due to the hermiticity of the Hamiltonian, one finds $d_{\mu\nu}^{\dagger}=d_{\mu\nu}$. With the help of the notations
\begin{gather}
    T_{i}=(d_{i1}\cos\phi+d_{i2}\sin\phi)\frac{\partial}{\partial r}+(-d_{i1}\sin\phi+d_{i2}\cos\phi)\frac{1}{r}\frac{\partial}{\partial \phi}-iEd_{i0},\\
    S_{i}=d_{i1}\mathcal{A}_{1}+d_{i2}\mathcal{A}_{2}=\frac{Br}{2}(d_{i1}\sin\phi-d_{i2}\cos\phi),
\end{gather}
the Dirac equation becomes
\begin{eqnarray}
          & & \left(E+m+K-\frac{i}{2}T_{3}-\frac{e}{2}S_{3}\right)\psi_{1}-\left[\frac{i}{2}(T_{1}-iT_{2})+\frac{e}{2}(S_{1}-iS_{2})\right]\psi_{2}- \nonumber \\
          & &\left(\frac{i}{2}T_{0}+\frac{e}{2}S_{0}\right)\psi_{3}+e^{-i\phi}\left(i\frac{\partial}{\partial r}+\frac{1}{r}\frac{\partial}{\partial \phi}+\frac{ieB r}{2}\right)\psi_{4}=0, 
\label{eq200} \\
        & &-\left[\frac{i}{2}(T_{1}+iT_{2})+\frac{e}{2}(S_{1}+iS_{2})\right]\psi_{1}+\left(E+m-K+\frac{i}{2}T_{3}+\frac{e}{2}S_{3}\right)\psi_{2}+ \nonumber \\
        & &e^{i\phi}\left(i\frac{\partial}{\partial r}-\frac{1}{r}\frac{\partial}{\partial \phi}-\frac{ieB r}{2}\right)\psi_{3}-\left(\frac{i}{2}T_{0}+\frac{e}{2}S_{0}\right)\psi_{4}=0,\label{eq201} \\
        & & \left(\frac{i}{2}T_{0}+\frac{e}{2}S_{0}\right)\psi_{1}-e^{-i\phi}\left(i\frac{\partial}{\partial r}+\frac{1}{r}\frac{\partial}{\partial \phi}+\frac{ieBr}{2}\right)\psi_{2}+ \nonumber \\
        & &\left(-E+m+K+\frac{i}{2}T_{3}+\frac{e}{2}S_{3}\right)\psi_{3}+\left[\frac{i}{2}(T_{1}-iT_{2})+\frac{e}{2}(S_{1}-iS_{2})\right]\psi_{4}=0, \label{eq202} \\
        &-& e^{i\phi}\left(i\frac{\partial}{\partial r}-\frac{1}{r}\frac{\partial}{\partial \phi}-\frac{i eBr}{2}\right)\psi_{1}+\left(\frac{i}{2}T_{0}+\frac{e}{2}S_{0}\right)\psi_{2} + \nonumber\\
        && \left[\frac{i}{2}(T_{1}+iT_{2})+\frac{e}{2}(S_{1}+iS_{2})\right]\psi_{3}+\left(-E+m-K-\frac{i}{2}T_{3}-\frac{e}{2}S_{3}\right)\psi_{4}=0. \label{eq203}
\end{eqnarray}

The contributions from $d_{30}$ can now be precisely solved. If we shut down all other $d_{\mu \nu}$'s, one can easily find the effects from $d_{30}$ can be attributed into the $\pm E + m \pm K$ terms. Similarly to (\ref{eq38}) and (\ref{eq39}), we have
\begin{equation}\label{eq282}
    \left(
      \begin{array}{cc}
        E+m+K-\frac{E}{2}d_{30}&i\left(\frac{n_{4}}{r_{c}}+\frac{eBr_{c}}{2}\right)\\
      i\left(\frac{n_{1}}{r_{c}}+\frac{eBr_{c}}{2}\right)&  -E+m-K-\frac{E}{2}d_{30}
      \end{array}
    \right)\left(
        \begin{array}{c}
            A_{1}\\
            A_{4}
        \end{array}
    \right)=0,
\end{equation}
\begin{equation}\label{eq283}
    \left(
      \begin{array}{cc}
        E+m-K+\frac{E}{2}d_{30}&-i\left(\frac{n_{3}}{r_{c}}+\frac{eBr_{c}}{2}\right)\\
       -i\left(\frac{n_{2}}{r_{c}}+\frac{eBr_{c}}{2}\right)& -E+m+K+\frac{E}{2}d_{30}
      \end{array}
    \right)\left(
        \begin{array}{c}
            A_{2}\\
            A_{3}
        \end{array}
    \right)=0
\end{equation}
and the solutions of which are
\begin{equation}\label{eq317}
    \frac{A_{1}}{A_{4}}=-\frac{i\left(\frac{n_{4}}{r_{c}}+\frac{p}{2}\right)}{E+m+K-\frac{E}{2}d_{30}},\ \frac{A_{4}}{A_{1}}=\frac{i\left(\frac{n_{1}}{r_{c}}+\frac{p}{2}\right)}{E-m+K+\frac{E}{2}d_{30}}
\end{equation}
\begin{equation}\label{eq318}
    \frac{A_{2}}{A_{3}}=\frac{i\left(\frac{n_{3}}{r_{c}}+\frac{p}{2}\right)}{E+m-K+\frac{E}{2}d_{30}},\ \frac{A_{3}}{A_{2}}=\frac{i\left(\frac{n_{2}}{r_{c}}+\frac{p}{2}\right)}{-E+m+K+\frac{E}{2}d_{30}}
\end{equation}
and
\begin{equation}
    n_{1}-n_{3}\simeq 2Kr_{c}+\frac{mr_{c}d_{30}E^{2}}{E^{2}-K^{2}}.
\end{equation}

Then we calculate the contributions from other $d_{\mu \nu}$'s except $d_{30}$ by perturbation theory. First we expand (\ref{GeneralPackets_Stationary}) perturbatively,
\begin{equation}\label{psiExpansion}
    \psi_{i}=\left\{A_{i}+\delta A_{i}+(B_{i}+\delta B_{i})\left[r-(r_{c}+\delta r_{c}(\phi))\right]^{2}\right\}e^{i(n_{i}+\delta n_{i}(\phi))\phi}e^{ik_{i}(\phi)r},
\end{equation}
and that all the non-zero components of $d_{\mu\nu}$ except $d_{30}$ are defined to be the same order as $\delta A_i$, $\delta B_i$, $\delta r_c$, $\delta n_i$ and $k_i$. Without loss of generality, $\delta A_i$, $\delta B_i$, $\delta r_c$, $k_i$ are assumed to be real and $\delta n_i$ can be complex. Therefore we can compare the terms order by order in the relatively complicated equations after inserting (\ref{psiExpansion}) into (\ref{eq200})-(\ref{eq203}). The zeroth order results are given by
\begin{equation}\label{eq216}
    \begin{aligned}
        &(E+m+K - \frac{E}{2} d_{30})A_{1}e^{in_{1}\phi}(i\delta n_{1}\phi+ik_{1}r_{c})-\frac{e}{2}S_{3}A_{1}e^{in_{1}\phi}-\frac{e}{2}(S_{1}-iS_{2})A_{2}e^{in_{2}\phi}-iA_{1}v_{3}n_{1}e^{in_{1}\phi}\\
       &-\frac{e}{2}S_{0}A_{3}e^{in_{3}\phi}-iA_{3}v_{0}n_{3}e^{in_{3}\phi}+\frac{i}{r_{c}}A_{4}(\delta n_{4}+\phi\partial_{\phi}\delta n_{4}+r_{c}\partial_{\phi}k_{4})e^{i(n_{4}-1)\phi}\\
       &-A_{2}v_{2}n_{2}e^{in_{2}\phi}-iA_{2}v_{1}n_{2}e^{in_{2}\phi}-k_{4}A_{4}e^{i(n_{4}-1)\phi}-\left(\frac{eB r_{c}}{2}+\frac{n_{4}}{r_{c}}\right)\left(\delta n_{4}\phi+k_{4}r_{c}\right)A_{4}e^{i(n_{4}-1)\phi}\\
       &
       +\left(i\frac{E}{2}d_{20}-\frac{E}{2}d_{10}\right)A_{2}e^{in_{2}\phi}-\frac{E}{2}d_{00}A_{3}e^{in_{3}}=0,
    \end{aligned}
\end{equation}
\begin{equation}\label{eq217}
    \begin{aligned}
        &(E+m-K + \frac{E}{2} d_{30})A_{2}e^{in_{2}\phi}(i\delta n_{2}\phi+ik_{2}r_{c})+\frac{e}{2}S_{3}A_{2}e^{in_{2}\phi}-\frac{e}{2}(S_{1}+iS_{2})A_{1}e^{in_{1}\phi}+iA_{2}v_{3}n_{2}e^{in_{2}\phi}\\
       &-\frac{e}{2}S_{0}A_{4}e^{in_{4}\phi}-iA_{4}v_{0}n_{4}e^{in_{4}\phi}-\frac{i}{r_{c}}A_{3}(\delta n_{3}+\phi\partial_{\phi}\delta n_{3}+r_{c}\partial_{\phi}k_{3})e^{i(n_{3}+1)\phi}\\
       &-A_{1}v_{2}n_{1}e^{in_{1}\phi}-iA_{1}v_{1}n_{1}e^{in_{1}\phi}-k_{3}A_{3}e^{i(n_{3}+1)\phi}+\left(\frac{eB r_{c}}{2}+\frac{n_{3}}{r_{c}}\right)(\delta n_{3}\phi+k_{3}r_{c})A_{3}e^{i(n_{3}+1)\phi}\\
       &
       -\left(\frac{E}{2}d_{10}+\frac{iE}{2}d_{20}\right)A_{1}e^{in_{1}\phi}-\frac{E}{2}d_{00}A_{4}e^{in_{4}\phi}=0,
    \end{aligned}
\end{equation}
\begin{equation}\label{eq218}
    \begin{aligned}
        &(-E+m+K + \frac{E}{2} d_{30})A_{3}e^{in_{3}\phi}(i\delta n_{3}\phi+ik_{3}r_{c})+\frac{e}{2}S_{3}A_{3}e^{in_{3}\phi}+\frac{e}{2}(S_{1}-iS_{2})A_{4}e^{in_{4}\phi}+iA_{3}v_{3}n_{3}e^{in_{3}\phi}\\
       &+\frac{e}{2}S_{0}A_{1}e^{in_{1}\phi}+iA_{1}v_{0}n_{1}e^{in_{1}\phi}-\frac{i}{r_{c}}A_{2}(\delta n_{2}+\phi\partial_{\phi}\delta n_{2}+r_{c}\partial_{\phi}k_{2})e^{i(n_{2}-1)\phi}\\
       &+A_{4}v_{2}n_{4}e^{in_{4}\phi}+iA_{4}v_{1}n_{4}e^{in_{4}\phi}+k_{2}A_{2}e^{i(n_{2}-1)\phi}+\left(\frac{eB r_{c}}{2}+\frac{n_{2}}{r_{c}}\right)(\delta n_{2}\phi+k_{2}r_{c})A_{2}e^{i(n_{2}-1)\phi}\\
       &
       +\left(\frac{E}{2}d_{10}-i\frac{E}{2}d_{20}\right)A_{4}e^{in_{4}\phi}+\frac{E}{2}d_{00}A_{1}e^{in_{1}\phi}=0,
    \end{aligned}
\end{equation}
\begin{equation}\label{eq219}
    \begin{aligned}
        &(-E+m-K - \frac{E}{2} d_{30})A_{4}e^{in_{4}\phi}(i\delta n_{4}\phi+ik_{4}r_{c})-\frac{e}{2}S_{3}A_{4}e^{in_{4}\phi}+\frac{e}{2}(S_{1}+iS_{2})A_{3}e^{in_{3}\phi}-iA_{4}v_{3}n_{4}e^{in_{4}\phi}\\
       &+\frac{e}{2}S_{0}A_{2}e^{in_{2}\phi}+iA_{2}v_{0}n_{2}e^{in_{2}\phi}+\frac{i}{r_{c}}A_{1}(\delta n_{1}+\phi\partial_{\phi}\delta n_{1}+r_{c}\partial_{\phi}k_{1})e^{i(n_{1}+1)\phi}\\
       &-A_{3}v_{2}n_{3}e^{in_{3}\phi}+iA_{3}v_{1}n_{3}e^{in_{3}\phi}+k_{1}A_{1}e^{i(n_{1}+1)\phi}-\left(\frac{eB r_{c}}{2}+\frac{n_{1}}{r_{c}}\right)(\delta n_{1}\phi+k_{1}r_{c})A_{1}e^{i(n_{1}+1)\phi}\\
       &
       +\left(\frac{E}{2}d_{10}+i\frac{E}{2}d_{20}\right)A_{3}e^{in_{3}\phi}+\frac{E}{2}d_{00}A_{2}e^{in_{2}\phi}=0,
    \end{aligned}
\end{equation}
while the first order terms are
\begin{equation}
    \begin{aligned}
        &ik_{1}(E+m+K-\frac{E}{2} d_{30})A_{1}e^{in_{1}\phi}+2iB_{4}e^{i(n_{4}-1)\phi}+\left(\frac{ieB}{2}-\frac{in_{4}}{r_{c}^{2}}\right)A_{4}e^{i(n_{4}-1)\phi}\\
        &-\frac{i}{r_{c}^{2}}\left(\delta n_{4}+\phi\partial_{\phi}\delta n_{4}+r_{c}\partial_{\phi}k_{4}\right)A_{4}e^{i(n_{4}-1)\phi}=0,
    \end{aligned}
\end{equation}
\begin{equation}
    \begin{aligned}
        &ik_{2}(E+m-K + \frac{E}{2} d_{30})A_{2}e^{in_{2}\phi}+2iB_{3}e^{i(n_{3}+1)\phi}-\left(\frac{ieB}{2}-\frac{in_{3}}{r_{c}^{2}}\right)A_{3}e^{i(n_{3}+1)\phi}\\
        &+\frac{i}{r_{c}^{2}}(\delta n_{3}+\phi\partial_{\phi}\delta n_{3}+r_{c}\partial_{\phi}k_{3})A_{3}e^{i(n_{3}+1)\phi}=0,
    \end{aligned}
\end{equation}
\begin{equation}
    \begin{aligned}
        &ik_{3}(-E+m+K + \frac{E}{2} d_{30})A_{3}e^{in_{3}\phi}-2iB_{2}e^{i(n_{2}-1)\phi}-\left(\frac{ieB}{2}-\frac{in_{2}}{r_{c}^{2}}\right)A_{2}e^{i(n_{2}-1)\phi}\\
        &+\frac{i}{r_{c}^{2}}(\delta n_{2}+\phi\partial_{\phi}\delta n_{2}+r_{c}\partial_{\phi}k_{2})A_{2}e^{i(n_{2}-1)\phi}=0,
    \end{aligned}
\end{equation}
\begin{equation}
    \begin{aligned}
        &ik_{4}(-E+m-K - \frac{E}{2} d_{30})A_{4}e^{in_{4}\phi}-2iB_{1}e^{i(n_{1}+1)\phi}+\left(\frac{ieB}{2}-\frac{in_{1}}{r_{c}^{2}}\right)A_{1}e^{i(n_{1}+1)\phi}\\
        &+\frac{i}{r_{c}^{2}}(\delta n_{1}+\phi\partial_{\phi}\delta n_{1}+r_{c}\partial_{\phi}k_{1})A_{1}e^{i(n_{1}+1)\phi}=0.
    \end{aligned}
\end{equation}
A critical point should be taken care with for further solving these equations. Because of the periodicity of the exponent in the wave function, the phase $\phi$ can be arbitrarily large, and thus the terms proportional to $\phi$ in (\ref{eq216})-(\ref{eq219}) are forced to disappear, leading us to the following constraints on $\delta n_i$,
\begin{equation}
    \begin{aligned}
        i\delta n_{1}(E+m+K - \frac{E d_{30}}{2})A_{1}e^{in_{1}\phi}+\frac{i}{r_{c}}A_{4}e^{i(n_{4}-1)\phi}\partial_{\phi}\delta n_{4}-\left(\frac{eB r_{c}}{2}+\frac{n_{4}}{r_{c}}\right)\delta n_{4}A_{4}e^{i(n_{4}-1)\phi}=0,
    \end{aligned}
\end{equation}
\begin{equation}
    \begin{aligned}
        i\delta n_{2}(E+m-K + \frac{E d_{30}}{2})A_{2}e^{in_{2}\phi}-\frac{i}{r_{c}}A_{3}e^{i(n_{3}+1)\phi}\partial_{\phi}\delta n_{3}+\left(\frac{eB r_{c}}{2}+\frac{n_{3}}{r_{c}}\right)\delta n_{3}A_{3}e^{i(n_{3}+1)\phi}=0,
    \end{aligned}
\end{equation}
\begin{equation}
    \begin{aligned}
        i\delta n_{3}(-E+m+K + \frac{E d_{30}}{2})A_{3}e^{in_{3}\phi}-\frac{i}{r_{c}}A_{2}e^{i(n_{2}-1)\phi}\partial_{\phi}\delta n_{2}+\left(\frac{eB r_{c}}{2}+\frac{n_{2}}{r_{c}}\right)\delta n_{2}A_{2}e^{i(n_{2}-1)\phi}=0,
    \end{aligned}
\end{equation}
\begin{equation}
    \begin{aligned}
        i\delta n_{4}(-E+m-K - \frac{E d_{30}}{2})A_{4}e^{in_{4}\phi}+\frac{i}{r_{c}}A_{1}e^{i(n_{1}+1)\phi}\partial_{\phi}\delta n_{1}-\left(\frac{eB r_{c}}{2}+\frac{n_{1}}{r_{c}}\right)\delta n_{1}A_{1}e^{i(n_{1}+1)\phi}=0.
    \end{aligned}
\end{equation}
These constraints can be further translated into (no summation on $j$)
\begin{equation}\label{eq238}
    \frac{i}{r_c}\partial_\phi \mathscr{N}_j=\mathscr{M}_j\mathscr{N}_j
\end{equation}
after taking (\ref{n1234}), (\ref{A4A1}) and (\ref{A3A2}) into account, where
\begin{equation}
    \mathscr{M}_1=\left(
        \begin{array}{cc}
             \frac{eBr_{c}}{2}+\frac{n_{4}}{r_{c}}&-\frac{n_{4}}{r_{c}}-\frac{eBr_{c}}{2}\\
             -\frac{n_{1}}{r_{c}}-\frac{eBr_{c}}{2}&\frac{eBr_{c}}{2}+\frac{n_{1}}{r_{c}}
        \end{array}
    \right),\ \ \mathscr{M}_2=\left(
        \begin{array}{cc}
             \frac{eBr_{c}}{2}+\frac{n_{3}}{r_{c}}&-\frac{n_{3}}{r_{c}}-\frac{eBr_{c}}{2}\\
             -\frac{n_{2}}{r_{c}}-\frac{eBr_{c}}{2}&\frac{eBr_{c}}{2}+\frac{n_{2}}{r_{c}}
        \end{array}
    \right)
\end{equation}
and $\mathscr{N}_1=\left(\delta n_4,\delta n_1\right)^\mathrm{T}$, $\mathscr{N}_2=\left(\delta n_3, \delta n_2\right)^\mathrm{T}$.
These equations can be solved by diagonalizing the $\mathcal{M}_{1,2}$ into
\begin{equation}
    \mathscr{U}_{j}^{-1} \mathscr{M}_j \mathscr{U}_{j}=\left(
        \begin{array}{cc}
            0&0\\
            0& \lambda_{j}
        \end{array}
    \right)
\end{equation}
with 
\begin{equation}
    \mathscr{U}_{1}=\left(
        \begin{array}{cc}
            1&-\frac{2n_{4}+eBr_{c}^{2}}{2n_{1}+eBr_{c}^{2}}\\
            1&1
        \end{array}
    \right),\mathscr{U}_{2}=\left(
        \begin{array}{cc}
            1&-\frac{2n_{3}+eBr_{c}^{2}}{2n_{2}+eBr_{c}^{2}}\\
            1&1
        \end{array}
    \right)
\end{equation}
and
\begin{equation}
\lambda_{1}=\frac{n_{1}+n_{4}+eB r_{c}^{2}}{r_{c}}, \quad \lambda_{2}=\frac{n_{2}+n_{3}+eBr_{c}^{2}}{r_{c}},
\end{equation}
(\ref{eq238}) will be transformed into
\begin{equation}
    \frac{i}{r_{c}}\partial_{\phi} \left(\mathscr{U}_{j}^{-1}\mathscr{N}_j\right)=\left(
        \begin{array}{cc}
            0&0\\
            0&\lambda_{j}
        \end{array}
    \right) \mathscr{U}_{j}^{-1}\mathscr{N}_j
\end{equation}
which can be readily solved to give 
\begin{gather}
    \delta n_{1}= \frac{C_{1}}{2n_{1}+2n_{4}+2pr_{c}}+\frac{2n_{1}+pr_{c}}{2n_{1}+2n_{4}+2pr_{c}}C_3e^{-i(n_{1}+n_{4}+pr_{c})\phi},\label{deltan1}\\
    \delta n_{2}=\frac{C_{2}}{2n_{2}+2n_{3}+2pr_{c}}+\frac{2n_{2}+pr_{c}}{2n_{2}+2n_{3}+2pr_{c}}C_4e^{-i(n_{2}+n_{3+pr_{c}})\phi},\\
    \delta n_{3}=\frac{C_{2}}{2n_{2}+2n_{3}+2pr_{c}}-\frac{2n_{3}+pr_{c}}{2n_{2}+2n_{3}+2pr_{c}}C_4e^{-i(n_{2}+n_{3+pr_{c}})\phi},\\
    \delta n_{4}=\frac{C_{1}}{2n_{1}+2n_{4}+2pr_{c}}-\frac{2n_{4}+pr_{c}}{2n_{1}+2n_{4}+2pr_{c}}C_3e^{-i(n_{1}+n_{4}+pr_{c})\phi}.\label{deltan4}
\end{gather}
To evaluate the precession frequency $w_p\propto n_{1}-n_{3}+\delta n_{1}-\delta n_{3}$, there are still several integral constants $C_{1,2,3,4}$ left to be determined. It is noteworthy that another constraint comes from the fact that the average radial Noether current vanishes since the muon wave-packet circulates within the uniform magnetic field as shown in Fig.~\ref{CyclotronSketch},
\begin{equation}\label{jmur}
    \lim_{\Phi \rightarrow \infty} \frac{2 \pi}{\Phi} \int_{0}^{\Phi}j^{\mu}\cdot rd\phi=0,
\end{equation}
where
\begin{equation}
    j^{\mu}=\overline{\psi}\gamma^{\mu}\psi+\frac{1}{2}d^{\nu\mu}\overline{\psi}\gamma_{5}\gamma_{\nu}\psi.
\end{equation}
It is straightforward to get
\begin{equation}
    \begin{aligned}
       &\lim_{\Phi \rightarrow \infty} \frac{2 \pi}{\Phi} \int_{0}^{\Phi}j^{\mu}\cdot rd\phi\\
       =&\lim_{\Phi \rightarrow \infty} \frac{2 \pi}{\Phi}  \int_0^{\Phi} d\phi \left\{\psi^{\dagger}\gamma^{0}\gamma^{1}\psi \cos\phi+\psi^{\dagger}\gamma^{0}\gamma^{2}\psi \sin\phi+\frac{d^{\nu 1}}{2}\psi^{\dagger}\gamma^{0}\gamma_{5}\gamma_{\nu}\psi \cos\phi+\frac{d^{\nu 2}}{2}\psi^{\dagger}\gamma^{0}\gamma_{5}\gamma_{\nu}\psi \sin\phi\right\}\\
       =&4\pi i\left[A_{1}^{\dagger}A_{4}(k_{4}-k_{1})r+A_{2}^{\dagger}A_{3}(k_{3}-k_{2})r\right]
    \end{aligned}.
\end{equation}
Here we note that all oscillation terms like $e^{i u \phi}$ with $u \neq 0$ vanishes, and we expand the results up to the lowest order. In the last step we have used the relations $A_{1}^{\dagger}A_{4}=-A_{4}^{\dagger}A_{1}$ and $A_{2}^{\dagger}A_{3}=-A_{3}^{\dagger}A_{2}$ from (\ref{A4A1}) and (\ref{A3A2}). Thus, the condition (\ref{jmur}) for any $A_1$ and $A_2$ configurations results in
\begin{equation}
        k_{1}=k_{4},\ k_{2}=k_{3}
\end{equation}
By Fourier expanding $k_i$ and $\delta r_c$,
\begin{equation}\label{foe}
\begin{aligned}
&k_{1}=k_{10}+k_{11}\cos[\delta n\phi]+k_{12}\sin[\delta n\phi]+k_{13}\cos[\theta_1\phi]+k_{14}\sin[\theta_1],\\
&k_{2}=k_{20}+k_{21}\cos[\delta n\phi]+k_{22}\sin[\delta n\phi]+k_{23}\cos[\theta_2\phi]+k_{24}\sin[\theta_2],\\
&k_{3}=k_{30}+k_{31}\cos[\delta n\phi]+k_{32}\sin[\delta n\phi]+k_{33}\cos[\theta_2\phi]+k_{34}\sin[\theta_2],\\
&k_{4}=k_{40}+k_{41}\cos[\delta n\phi]+k_{42}\sin[\delta n\phi]+k_{43}\cos[\theta_1\phi]+k_{44}\sin[\theta_1],\\
& \delta r_{c}=r_{0}+r_{1}\cos[\delta n\phi]+s_{1}\sin[\delta n\phi]+r_{2}\cos[(\theta_1\phi]+s_{2}\sin[\theta_1\phi]+r_{3}\cos[\theta_2\phi]+s_{3}\sin[\theta_2\phi]
\end{aligned}
\end{equation}

with $\delta n\equiv n_1-n_3$, $\theta_1\equiv n_1+n_4+pr_c$ and $\theta_2\equiv n_2+n_3+pr_c$, one can see that all the Fourier coefficients of the trigonometric functions should be zero except for the zero order one, i.e., 
\begin{equation}\label{k14k23}
 k_{10}=k_{40}\equiv k_0,\quad k_{20}=k_{30}\equiv k'_0,  \quad \delta r_c=r_0,
\end{equation}
which simplifies (\ref{eq216})-(\ref{eq219}) to 
\begin{equation}
    \begin{aligned}
        i(E+K+m - \frac{E d_{30}}{2})k_{0}r_{c}A_{1}-k_{0}A_{4}\left(1+n_{4}+\frac{pr_{c}}{2}\right)+\frac{iC_{1}A_{4}}{2r_{c}\theta_1}=0,
    \end{aligned}
\end{equation}
\begin{equation}
    \begin{aligned}
        i(E-K+m + \frac{E d_{30}}{2})k'_{0}r_{c}A_{2}+k'_{0}A_{3}\left(-1+n_{3}+\frac{pr_{c}}{2}\right)-\frac{iC_{2}A_{3}}{2r_{c}\theta_2}=0,
    \end{aligned}
\end{equation}
\begin{equation}
    \begin{aligned}
        i(-E+K+m + \frac{E d_{30}}{2})k'_{0}r_{c}A_{3}+k'_{0}A_{2}\left(1+n_{2}+\frac{pr_{c}}{2}\right)-\frac{iC_{2}A_{2}}{2r_{c}\theta_2}=0,
    \end{aligned}
\end{equation}
\begin{equation}
    \begin{aligned}
        i(-E-K+m - \frac{E d_{30}}{2})k_{0}r_{c}A_{4}+k_{0}A_{1}\left(1-n_{1}-\frac{pr_{c}}{2}\right)+\frac{iC_{1}A_{1}}{2r_{c}\theta_1}=0,
    \end{aligned}
\end{equation}
Utilizing (\ref{eq317}) and (\ref{eq318}) these can be further simplified to
\begin{gather}
    -k_{0}+\frac{iC_{1}}{2r_{c}\theta_1}=0,\label{eq313}\\
    -k'_{0}-\frac{iC_{2}}{2r_{c}\theta_2}=0,\label{eq314}\\
    k'_{0}-\frac{iC_{2}}{2r_{c}\theta_2}=0,\label{eq315}\\
    k_{0}+\frac{iC_{1}}{2r_{c}\theta_1}=0,\label{eq316}
\end{gather}
whose solutions are
\begin{equation}\label{c1c20}
    C_1=C_2=0.
\end{equation}
Finally, neglecting the extremely rapidly oscillating terms in (\ref{deltan1})-(\ref{deltan4}), whose averaged contributions to the precession processes in a macroscopic period of time should vanish, we arrive at 
\begin{equation}
    \begin{aligned}
         \delta n_{3}-\delta n_{1}&=\frac{C_{2}}{2(n_{2}+n_{3}+pr_{c})}-\frac{C_{1}}{2(n_{1}+n_{4}+pr_{c})}=0.
    \end{aligned}
\end{equation}

This result means that the first order contribution $\delta n_1-\delta n_3$ vanishes and the precession frequency is 
\begin{equation}\label{resultd2}
    \omega_a = \left\lbrace \begin{array}{l} (n_{3}-n_{1}  +\delta n_{3}-\delta n_{1})\frac{eB}{|E|}\text{, for $E>0$} \\ (n_{1}-n_{3} + \delta n_{1}-\delta n_{3})\frac{eB}{|E|}\text{, for $E<0$} \end{array} \right. \simeq -2K\frac{eBr_{c}}{|E|}-\frac{md_{30}E^{2}}{E^{2}-K^{2}}\frac{eBr_{c}}{|E|}.
\end{equation}
To summarize this section, the contribution of $d_{\mu\nu}$ term on $\omega_a$ are characterized by \eqref{resultd2}.

\section{contribution of other Lorentz violating terms}
Utilizing the perturbation method and similar techniques presented in the previous section, we comprehensively study the contribution of all the other Lorentz violating terms in (\ref{action}), namely $-a_{\mu}\overline{\psi}\gamma^{\mu}\psi$, $-b_{\mu}\overline{\psi}\gamma_{5}\gamma^{\mu}\psi$, $\frac{i}{2}c_{\mu\nu}\overline{\psi}\gamma^{\mu}D^{\nu}\psi$ and $-\frac{1}{2}H_{\mu\nu}\overline{\psi}\sigma^{\mu\nu}\psi$,
on the precession frequency and gather the results below, leaving the more technical details in Appendix. \ref{app_o}.

For $a_\mu$, we found that the solution of a non-zero $a_0$ does not require perturbation theory, so we acquire a precise solution when $a_1$, $a_2$ and $a_3$ vanish,
\begin{equation}\label{resulta}
    \omega_{a} =\left\lbrace \begin{array}{l} (n_{3}-n_{1}  +\delta n_{3}-\delta n_{1})\frac{eB}{|E|}\text{, for $E>0$} \\ (n_{1}-n_{3} + \delta n_{1}-\delta n_{3})\frac{eB}{|E|}\text{, for $E<0$} \end{array} \right.= - \left(2K+\frac{Km^{2}}{(E+a_{0})^{2}-K^{2}}\right)\frac{eBr_{c}}{|E|}.
\end{equation}
The dependence on $a_{1,2,3}$ is calculated through perturbation expansions. Their leading order contributions are zero. In fact, expanding (\ref{resulta}) into series on $a_0$ gives the leading contribution $\propto K a_0$, with its effect suppressed by a factor of $\frac{K}{E}$, thus is compatible with the null leading order results addressed in Ref.~\cite{Bluhm:1999dx, Bluhm:1997qb, Bluhm:1997ci}.

For $b_\mu$, the $b_3$ contributions do not require perturbative expansions,
\begin{equation}\label{resultb}
    \begin{aligned}
        \omega_{a}& =\left\lbrace \begin{array}{l} (n_{3}-n_{1}  +\delta n_{3}-\delta n_{1})\frac{eB}{|E|}\text{, for $E>0$} \\ (n_{1}-n_{3} + \delta n_{1}-\delta n_{3})\frac{eB}{|E|}\text{, for $E<0$} \end{array} \right.=-\left(2K-\frac{2mb_{3}E-K(m^{2}+b_{3}^{2})}{E^{2}-K^{2}}\right)\frac{eBr_{c}}{E}\\
        &=-\left(2K\pm\frac{2mb_{3}|E|}{E^{2}-K^{2}}\right)\frac{eBr_{c}}{|E|},
    \end{aligned} 
\end{equation}
where we note ``$+$'' is adopted for $E>0$ and ``$-$'' is adopted for $E<0$. This induces a difference between particle and anti-particle, which is a CPT-breaking effect, as illustrated by Ref.~\cite{Bluhm:1999dx, Bluhm:1997qb, Bluhm:1997ci, Kostelecky:2013rta}. The $b_0$, $b_1$ and $b_2$ contributions again vanish up to their leading perturbative contributions.

For $c_{\mu\nu}$, the contributions can be calculated
\begin{equation}\label{reslutc1}
 \omega_a=(n_1-n_3+\delta n_1-\delta n_3)\frac{eB}{E}\simeq \frac{2Kp}{E}+\frac{(c_{11}+c_{22})eB}{4E}
\end{equation}
It is compatible with the result addressed in Ref.~\cite{Crivellin:2022idw}.


For $H_{\mu\nu}$, again the contribution from $H'_{12}$ can be calculated precisely, where $H'_{\mu\nu}\equiv2H_{[\mu\nu]}=H_{\mu\nu}-H_{\nu\mu}$, so we find
\begin{equation}\label{resultH}
    \omega_{a}=\left\lbrace\begin{array}{l} (n_{3}-n_{1}  +\delta n_{3}-\delta n_{1})\frac{eB}{|E|}\text{, for $E>0$} \\ (n_{1}-n_{3} + \delta n_{1}-\delta n_{3})\frac{eB}{|E|}\text{, for $E<0$} \end{array} \right.\simeq 2(K+H'_{12})\frac{eBr_{c}}{|E|}\left(1+\frac{m^{2}}{E^{2}-(K+H'_{12})^{2}}\right).
\end{equation}
Effects from other terms are again calculated perturbatively and their leading order contributions vanish.

In conclusion, \eqref{omegap},  \eqref{resultd2}, \eqref{resulta}, \eqref{resultb}, 
\eqref{reslutc1},
and  \eqref{resultH} consist our results for the contribution of all the terms in \eqref{action} to the precession frequency of muon.

\section{Summary and some additional instructions}

In this paper, we rely on the (modified) Dirac equations to resolve an old problem to calculate the precession frequency for a muon particle circulating within a uniform magnetic field as a simplified ``mind experiment''. We achieve similar results compared with the literature, and further show the existence of the tiny quantum corrections in (\ref{QuantumCorrections}). The particles in this paper are regarded as wave packets, and the tactic described in this paper accommodate the potential to be improved towards further perturbative expansions. We would also like to note that practical experiments involve both electric and magnetic fields configured within the equipment to stabilize the muon particle's status. Introducing these complexities requires a modification on (\ref{VectorPotential}), and most of the calculation processes can be easily corrected.

\begin{acknowledgements}
We thank to Chengfeng Cai for helpful discussions. We appreciate Fiona Kirk on behalf of Andreas Crivellin and Marco Schreck for his valuable communication. This work is supported in part by the National Natural Science Foundation of China under Grants Nos. 12005312, 12275367,
the China Postdoctoral Science Foundation under Grants No. 2022M723677, the Guangzhou Science and Technology Program under Grant No.202201011556, the Fundamental Research Funds for the Central
Universities, the Natural Science Foundation of Guangdong Province, and the Sun Yat-Sen University Science Foundation. 
\end{acknowledgements}

\appendix
\section{the calculation for other Lorentz violating terms}\label{app_o}
In this appendix we do not wish to disturb the readers with similar cumbersome computation, but instead  present some key expressions during the calculation of the contributions from other Lorentz violating couplings, $a_\mu$, $b_\mu$, $c_{\mu\nu}$ and $H_{\mu\nu}$.

In the case of $a_\mu$, (\ref{eq38}) and (\ref{eq39}) accordingly turn into
\begin{equation}
    \left(
        \begin{array}{cc}
            E+m+K+a_{0}&\frac{in_{4}}{r_{c}}+\frac{ieBr_{c}}{2}\\
            \frac{ieBr_{c}}{2}+\frac{in_{1}}{r_{c}}&-E+m-K-a_{0}
        \end{array}
    \right)\left(
        \begin{array}{c}
            A_{1}\\
            A_{4}
        \end{array}
    \right)=0
\end{equation}
\begin{equation}
    \left(
        \begin{array}{cc}
            E+m-K+a_{0}&-\frac{in_{3}}{r_{c}}-\frac{ieBr_{c}}{2}\\
            -\frac{ieBr_{c}}{2}-\frac{in_{2}}{r_{c}}&-E+m+K-a_{0}
        \end{array}
    \right)\left(
        \begin{array}{c}
            A_{2}\\
            A_{3}
        \end{array}
    \right)=0,
\end{equation}
whose solutions give
\begin{gather}
    \frac{A_{1}}{A_{4}}=\frac{-i\left(\frac{n_{4}}{r_{c}}+\frac{p}{2}\right)}{E+m+K+a_{0}},\ \frac{A_{4}}{A_{1}}=\frac{i\left(\frac{n_{1}}{r_{c}}+\frac{p}{2}\right)}{E-m+K+a_{0}},\\
    \frac{A_{2}}{A_{3}}=\frac{i\left(\frac{n_{3}}{r_{c}}+\frac{p}{2}\right)}{E+m-K+a_{0}},\ \frac{A_{3}}{A_{2}}=\frac{i\left(\frac{n_{2}}{r_{c}}+\frac{p}{2}\right)}{-E+m+K-a_{0}},
\end{gather}
and
\begin{equation}
    n_1-n_3=2Kr_c+\frac{Km^{2}r_c}{(E+a_{0})^{2}-K^{2}},
\end{equation}
similar to the case of $d_{\mu\nu}$ as in \eqref{eq282} and \eqref{eq283}.
Again, the constrains via the disappearance of the terms $\propto\phi$ still take the same form as (\ref{eq238}),
\begin{equation}
    \frac{i}{r_c}\partial_\phi \mathscr{N}_j=\mathscr{M}_j\mathscr{N}_j
\end{equation}
with
\begin{equation}
    \mathscr{M}_1=\left(
        \begin{array}{cc}
             \frac{eBr_{c}}{2}+\frac{n_{4}}{r_{c}}&-\frac{n_{4}}{r_{c}}-\frac{eBr_{c}}{2}\\
             -\frac{n_{1}}{r_{c}}-\frac{eBr_{c}}{2}&\frac{eBr_{c}}{2}+\frac{n_{1}}{r_{c}}
        \end{array}
    \right),\ \ \mathscr{M}_2=\left(
        \begin{array}{cc}
             \frac{eBr_{c}}{2}+\frac{n_{3}}{r_{c}}&-\frac{n_{3}}{r_{c}}-\frac{eBr_{c}}{2}\\
             -\frac{n_{2}}{r_{c}}-\frac{eBr_{c}}{2}&\frac{eBr_{c}}{2}+\frac{n_{2}}{r_{c}}
        \end{array}
    \right)
\end{equation}
and $\mathscr{N}_1=\left(\delta n_4,\delta n_1\right)^\mathrm{T}$, $\mathscr{N}_2=\left(\delta n_3, \delta n_2\right)^\mathrm{T}$.
Thus the similarity transformation afterwards leads to the  solutions of $\delta n_i$ being formally identical to (\ref{deltan1})-(\ref{deltan4}). Furthermore, the deduced equations from the constraint of vanishing radial Noether current are identical to \eqref{eq313}-\eqref{eq316}.
Consequently, we obtain the result  \eqref{resulta}.

Analogous calculation and   discussion can be applied in the cases of $b_\mu$ and $H_{\mu\nu}$ to arrive at the results \eqref{resultb} and \eqref{resultH}, while in the former case the analogy of (\ref{eq38}) and (\ref{eq39}) are 
\begin{equation}
    \left(
        \begin{array}{cc}
            E+m+K+b_{3}&\frac{ieBr_{c}}{2}+\frac{in_{4}}{r_{c}}\\
            \frac{ieBr_{c}}{2}+\frac{in_{1}}{r_{c}}&-E+m-K+b_{3}
        \end{array}
    \right)\left(
        \begin{array}{c}
            A_{1}\\
            A_{4}
        \end{array}
    \right)=0,
\end{equation}
\begin{equation}
    \left(
        \begin{array}{cc}
            E+m-K-b_{3}&-\left(\frac{ieBr_{c}}{2}+\frac{in_{3}}{r_{c}}\right)\\
            -\left(\frac{ieBr_{c}}{2}+\frac{in_{2}}{r_{c}}\right)&-E+m+K-b_{3}
        \end{array}
    \right)\left(
        \begin{array}{c}
            A_{2}\\
            A_{3}
        \end{array}
    \right)=0,
\end{equation}
and in the latter 
\begin{equation}
    \left(
        \begin{array}{cc}
            E+m+K+H'_{12}&\frac{ieBr_{c}}{2}+\frac{in_{4}}{r_{c}}\\
            \frac{ieBr_{c}}{2}+\frac{in_{1}}{r_{c}}&-E+m-K-H'_{12}
        \end{array}
    \right)\left(
        \begin{array}{c}
             A_{1}\\
        A_{4}
        \end{array}  
    \right)=0,
\end{equation}
\begin{equation}
    \left(
        \begin{array}{cc}
            E+m-K-H'_{12}&-\frac{ieBr_{c}}{2}-\frac{in_{3}}{r_{c}}\\
            -\frac{ieBr_{c}}{2}-\frac{in_{2}}{r_{c}}&-E+m+K+H'_{12}
        \end{array}
    \right)\left(
        \begin{array}{c}
             A_{2}\\
        A_{3}
        \end{array}  
    \right)=0,
\end{equation}
where $H'_{12}\equiv H_{12}-H_{21}$.

For the $c_{\mu \nu}$, $c_{00}$ can be attributed to the zeroth order, so the corresponding equations are
\begin{equation}
    \left(
      \begin{array}{cc}
        E+m+K+\frac{Ec_{00}}{2}&i\left(\frac{n_{4}}{r_{c}}+\frac{eBr_{c}}{2}\right)\\
      i\left(\frac{n_{1}}{r_{c}}+\frac{eBr_{c}}{2}\right)&  -E+m-K-\frac{Ec_{00}}{2}
      \end{array}
    \right)\left(
        \begin{array}{c}
            A_{1}\\
            A_{4}
        \end{array}
    \right)=0,
\end{equation}
\begin{equation}
    \left(
      \begin{array}{cc}
        E+m-K+\frac{Ec_{00}}{2}&-i\left(\frac{n_{3}}{r_{c}}+\frac{eBr_{c}}{2}\right)\\
       -i\left(\frac{n_{2}}{r_{c}}+\frac{eBr_{c}}{2}\right)& -E+m+K-\frac{Ec_{00}}{2}
      \end{array}
    \right)\left(
        \begin{array}{c}
            A_{2}\\
            A_{3}
        \end{array}
    \right)=0,
\end{equation}
The $c_{00}$ terms only translate the $E$, therefore no additional contributions to $n_{1}-n_{3}$ arise so it is the same with (\ref{n3-n1}), and the constrains via the disappearance of the terms $\propto\phi$ still leads to the results  (\ref{deltan1})-(\ref{deltan4}). However,
the analogy of \eqref{jmur} turns to be 
\begin{eqnarray}
     &&\frac{1}{2n\pi} \int_{0}^{2n\pi}j^{\mu}\cdot \overrightarrow{r}d\phi = 2i(\frac{c_{11}+c_{22}}{4}+1)[A_{1}^{\dagger}A_{4}(k_{4}-k_{1})r+A_{2}^{\dagger}A_{3}(k_{3}-k_{2})r] \nonumber \\
     &&+ \frac{i(c_{12}-c_{21})}{2}(A_{1}^{\dagger}A_{4}-A_{2}^{\dagger}A_{3}). \label{RadiusNoether}
\end{eqnarray}
There is no simple relations among $k_{1,2,3,4}$ to eliminate (\ref{RadiusNoether}). Therefore Fourier expanding in the same form as \eqref{foe} can be performed while condition \eqref{k14k23} should be replaced with
\begin{equation}
    2i(\frac{c_{11}+c_{22}}{4}+1)[A_{1}^{\dagger}A_{4}(k_{40}-k_{10})r+A_{2}^{\dagger}A_{3}(k_{30}-k_{20})r] + \frac{i(c_{12}-c_{21})}{2}(A_{1}^{\dagger}A_{4}-A_{2}^{\dagger}A_{3}) = 0
    \label{ConstraintJ}
\end{equation}
then the zero order contribution is same as the one for $\mathcal{A}_\mu$, while the first order contribution are as followings:
\begin{equation}\label{eqA12}
    \begin{aligned}
        \frac{iC_{1}}{2r_{c}\theta_{1}}+(n_{4}+\frac{pr_{c}}{2})k_{10}-(n_{4}+\frac{pr_{c}}{2}+1)k_{20}+\frac{1}{2}(\frac{n_{4}}{2r_{c}}+\frac{p}{4})(-c_{12}+ic_{22}+ic_{11}+c_{21}) &=& 0\\
      \frac{iC_{1}}{2r_{c}\theta_{1}}+(n_{1}+\frac{pr_{c}}{2})k_{20}-(n_{1}+\frac{pr_{c}}{2}-1)k_{10}-\frac{1}{2}(\frac{n_{1}}{2r_{c}}+\frac{p}{4})(c_{12}+ic_{22}+ic_{11}-c_{21}) &=& 0\\  
       -\frac{iC_{2}}{2r_{c}\theta_{2}}-(n_{3}+\frac{pr_{c}}{2})k_{30}+(n_{3}+\frac{pr_{c}}{2}-1)k_{40}-\frac{1}{2}(\frac{n_{3}}{2r_{c}}+\frac{p}{4})(c_{12}+ic_{22}+ic_{11}-c_{21}) &=& 0\\
      -\frac{iC_{2}}{2r_{c}\theta_{2}}-(n_{2}+\frac{pr_{c}}{2})k_{40}+(n_{2}+\frac{pr_{c}}{2}+1)k_{30}+\frac{1}{2}(\frac{n_{2}}{2r_{c}}+\frac{p}{4})(-c_{12}+ic_{22}+ic_{11}+c_{21}) &=& 0
  \end{aligned}
\end{equation}
The number of the unknown parameters are more than the number of equations. eliminating as many of the unknown numbers as possible, we acquire
\begin{equation}
    \begin{aligned}
            \delta n_{1}-\delta n_{3}&=\frac{C_{1}}{2(n_{1}+n_{4}+pr_{c})}-\frac{C_{2}}{2(n_{2}+n_{3}+pr_{c})}\\
            &=\frac{ipr_{c}}{2}(c_{21}-c_{12})+\frac{i}{4}(c_{21}-c_{12}+ic_{11}+ic_{22})+\frac{i}{p}(k_{10}-k_{20})\\
            &+\frac{c_{11}+c_{22}}{2}-\frac{i}{4pr_{c}}(c_{21}-c_{12}+ic_{11}+ic_{22})
    \end{aligned}
\end{equation}
 Due to the imaginary part of the $\delta n_{1}-\delta n_{3}$ will lead to the unphysical exponential decay or increase of the wave function as $\phi$ varies without contributing to the precession, it must be set zero by proper choices of $k_{10}$ and $k_{20}$. Even combined with the (\ref{ConstraintJ}) which is linear on all $k_{10, 20, 30, 40}$, this is still easy to accomplish. Finally because of $c_{\mu\nu}^{*}=c_{\mu\nu}$, we can obtain the real part of $\delta n_{1}-\delta n_{3}$ which affects the precession frequency:
 \begin{equation}\label{eqA16}
    Re(\delta n_{1}-\delta n_{3})=\frac{1}{4}(c_{11}+c_{22})+\frac{1}{4pr_{c}}(c_{11}+c_{22})\simeq \frac{c_{11}+c_{22}}{4}
\end{equation}

which leads to $n_{1}-n_{3}\simeq 2Kr_{c}$.
\newpage

\bibliographystyle{utphys}
\bibliography{ref}

\end{document}